\documentclass[preprint,12pt]{elsarticle}
\usepackage{amssymb,graphicx}

\journal{}

\def\be{\begin{equation}}
\def\ee{\end{equation}}
\def\bea{\begin{eqnarray}}
\def\eea{\end{eqnarray}}

\def\bL{L^\dag}

\def\cd{{\cal D}}

\def\cd{{\cal D}}

\def\pd{\partial}
\def\d{\mathrm{d}}

\def\exp{\mathrm{exp}}
\def\ex{\mathrm{e}}

\def\etal{\textit{et.al.}}
\def\ie{\textit{i.e.} }

\begin{document}

\begin{frontmatter}
\title{The thermodynamic properties of Davydov-Scott's  protein model in thermal bath}

\author{A. Sulaiman$^{a,b,d}$, F.P Zen$^{a,d}$, H. Alatas$^{c,d}$ and L.T. Handoko$^{e,f}$}

\address{$^{a)}$Theoretical Physics Laboratory, THEPI Research
Division, Institut Teknologi Bandung, Jl. Ganesha 10, Bandung
40132, Indonesia}

\address{ $^{b)}$Badan Pengkajian dan Penerapan Teknologi, BPPT Bld.
II (19$^{\rm th}$ floor), Jl. M.H. Thamrin 8, Jakarta 10340,
Indonesia\\
Email : asulaiman@webmail.bpp.go.id}

\address{ $^{c)}$Theoretical Physics Division, Department of
Physics, Bogor Agricultural University, Kampus Dermaga, Bogor,
Indonesia  }

\address{$^{d)}$Indonesia Center for Theoretical and Mathematical
Physics (ICTMP), Jl. Ganesha 10, Bandung 40132, Indonesia}

\address{$^{e)}$Group for Theoretical and Computational Physics, Research
Center for Physics, Indonesian Institute of Sciences, Kompleks
Puspiptek Serpong, Tangerang 15310, Indonesia}

\address{$^{f)}$Department of Physics, University of Indonesia,
Kampus UI Depok, Depok 16424, Indonesia }

\begin{abstract}
The thermodynamic properties of Davydov-Scott monomer contacting with thermal bath is investigated using Lindblad open quantum system formalism. The Lindblad equation is investigated through path integral method. It is found that the environmental effects contribute destructively to the specific heat, and large interaction between amide-I and amide-site is not preferred for a stable Davydov-Scott monomer.
\end{abstract}

\begin{keyword}
Davydov-Scott \sep open quantum system \sep specific heat
\end{keyword}

\end{frontmatter}

\section{Introduction}
\label{sec:intro}

Does the Davydov-Scott's soliton exist at biological temperature ?
The question has attracted more interest in the last  decades
\cite{cruzeiro1, cruzeiro2, scott, cuevas}. Earlier studies using
finite temperature molecular dynamics showed that the Davydov
soliton lifetime is only few picoseconds which is too short at the
biological temperature. The reason is the random thermal prevents
Davydov self-trapping from occurring as, for example as discussed
in \cite{hanson} which showed that the two-quantum state might be
more stable than the one-quantum state. Furthermore, using the
standard Davydov model, some numerical calculations also indicated
that soliton is stable at 310K. On the other hand, the analytic
calculation based on trial function or perturbation methods
obtained that soliton is stable at 300K \cite{cottingham, kapor1}.

The above-mentioned calculations were performed using the equilibrium quantum
system at finite temperature \cite{luzzia}. The finite temperature means that the quantum
system is in contact with environment such as thermal bath. The interaction of
a system with its environment is given by the dissipation effect
in quantum system. However, the dissipation effect
leads to a serious problem for quantization procedure. The most
appropriate theory to resolve this problem is the Quantum State
Diffusion (QSD) based on Lindblad formulation \cite{perceval}. The
first application of QSD to the protein model has been done by
Cuevas \etal \cite{cuevas}. Their calculation on Davydov-Scott
monomer showed that at room temperature the semi classical
approach might be a good approximation compared to the
corresponding full quantum system. However the study was focused
on the dynamical aspect of the system, \ie the solution of
Heisenberg's picture and its wave function based on the QSD equation. Recent
studies of the anharmonic effect for the monomer has also been
done by us using path integral and the thermodynamic function \cite{sulaiman}.
The advantage of calculating thermodynamic function is also relevant in
another approaches such as the models describing the phenomena in term of
elementary matter interactions using lagrangian \cite{sulaiman2}.
It was argued that the thermodynamic properties should be easier to observe
than another ones based on the wave function. Therefore, it is important to
study such system from statistical mechanics point of view.

The paper studies the thermodynamic properties of Davydov-Scott monomer including the thermal bath. The effect of thermal bath is investigated using the Lindblad open quantum system formalism through path integral method. The temperature dependencies of the system are obtained for the normalized specific heat. It is shown that the specific heat is sensitive to the size of coupling constant between amide-I and amide-site.

The paper is organized as follows. In  Sec. \ref{sec:model}  the
Davydov Hamiltonian for a one-dimensional molecular monomer is
described and the thermal bath effect is studied in Lindblad
equation.  In Sec. \ref{sec:tp} the thermodynamic properties are
investigated using path integral method. The paper is ended by a short summary.

\section{Lindblad open quantum system formulation}
\label{sec:model}

We use Davydov-Scott's model of the alpha-helix protein. The
Davydov-Scott monomer is a coupled of the amide-I oscillator that
expressed by the coordinate ($x$) and momentum ($p$) operators,
and the amide-site is expressed by the displacement and
momentum operators, $Q$ and $P$, respectively. Hamiltonian in the model has
the form \cite{sulaiman},
\be
   H = \frac{p^2}{2m} + \frac{1}{2} m \omega^2 x^2 +\frac{P^2}{2M}+
\frac{1}{2}\kappa Q^2 + \chi x Q \;
   ,
   \label{eq:mono1}
\ee
where $\omega$ is the intrinsic frequency of amide-I oscillation, $\chi$ is
the coupling  constant between two oscillators, $m$ ($M$) is the amide-I
(amide-site) mass and $\delta$ is the anharmonic coefficient.
Throughout the paper we also use the notations
$\chi'=\chi\sqrt{{2M\Omega}/\hbar}$ and $\Omega=\sqrt{\kappa/M}$. The
Hamiltonian describes the Davydov-Scott's monomer as a coupled harmonic
oscillator.

If the environmental or dissipation effect can not be ignored, the physical
system is not reversible. In another words, the irreversibility implies the
dissipation effect in a quantum system under consideration. One of the
basic tools to introduce dissipation in quantum mechanics is the
dynamical semi groups, and called as the Lindblad open quantum system formalism.
In particular, the quantum system whose the wave function equation can be
obtained from the Lindblad equation is called as QSD \cite{perceval}.

According to Lindblad formalism, the usual von Neumann-Liouville equation is
replaced by Lindblad equation or master equation in the form of,
\be
 \frac{\pd \rho}{\pd
 t}=-\frac{i}{\hbar}[H,\rho]+\sum_{j}(L_j\rho\bL_j
 -\frac{1}{2}\bL_j L_j \rho - \frac{1}{2} \rho \bL_j L_j) \; .\
 \label{eq:master1}
\ee
$\rho$ is the density function and $L_j$ denotes the Lindblad operator which
may neither Hermitian nor unique. In this formalism, the operator $H$ describes
internal dynamics, while $L$ represents the environmental effects
in the system.

Throughout the paper, let us assume that the environmental strength of the
amide-site is stronger than the amide-I excitation. Since $L$ must be the first
order in $Q$ and $P$, we choose the Lindblad operators as follow,
\bea
  L_1 &=& \sqrt{\gamma(1+\nu)} \left( \sqrt{\frac{M \Omega}{2\hbar}}Q+i\sqrt{\frac{1}{2M\hbar \Omega}}P
  + \frac{\chi'}{\hbar \Omega} \sqrt{\frac{m \omega}{2\hbar}}x \right) \; ,
  \label{eq:operator1}\\
  L_2 &=& \sqrt{\gamma\nu} \left( \sqrt{\frac{M\Omega}{2\hbar}}Q-i\sqrt{\frac{1}{2M \hbar \Omega}}P
  +  \frac{\chi'}{\hbar \Omega} \sqrt{\frac{m \omega}{2\hbar}}x \right)  \; .\
 \label{eq:operator2}
\eea
$\gamma$ is a damping parameter and $\nu= \left(
\ex^{{\hbar\Omega}/{k_BT}}-1\right)^{-1}$ is Bose-Einstein distribution
function with the Boltzman coefficient $k_B$.

The master equation in Eq.(\ref{eq:master1}) is calculated using
 Feynman path integral. Assuming the diffusion term is dominant
over the frictional damping rate ($[Q,[Q,\rho]] \gg [Q,[P,\rho]]$),
Eq. (\ref{eq:master1}) is rewritten in a differential
representation,
\bea
   \frac{\pd \rho}{\pd t} & = & \frac{i\hbar}{2m}
   \left( \frac{\pd^2}{\pd x^2}-\frac{\pd^2}{\pd x'^2} \right) \rho -
   \left(\frac{ im\omega^2}{2\hbar}+\frac{\delta_3}{2\hbar^2}\right) \left( x^2 - x'^2 \right) \rho
         \nonumber \\
   &&+ \left(\frac{i\hbar}{2M}+\frac{\delta_1}{2\hbar}\right) \left(
\frac{\pd^2}{\pd Q^2} - \frac{\pd^2}{\pd Q'^2} \right) \rho
    - \left(\frac{i M \Omega^2 }{2\hbar}+\frac{
    \delta_2}{2\hbar}\right)(Q^2-Q^{'2})\rho \nonumber\\
    &&- i \left( \frac{ \chi}{\hbar}+\frac{\delta_4}{2\hbar} \right) \left( xQ-
x'Q' \right) \rho \; .
  \label{eq:pdemaster}
\eea
Here, the Lindblad's coefficients are $\delta_1= \gamma(1+2\nu)/(2M\Omega)$,
$\delta_2 = {\gamma (1+2 \nu) M\Omega}/2$,
$\delta_3 = \gamma(1+2\nu){(\chi^2m\omega)}/(\hbar \Omega)^2$ and
$\delta_4 = \sqrt{m \omega M \Omega} \chi'/{(\hbar \Omega)} $.
The propagator of Eq. (\ref{eq:pdemaster}) is given by,
\bea
 K(x,x';Q,Q') & = & \int \int \cd [x] \cd [Q]
           \exp \left[ \frac{i}{\hbar}\int \d t \left( \frac{1}{2} m \dot{x}^2 - \frac{1}{2}m \tilde{\omega}^2 x^2
         \right. \right. \nonumber \\
       &&+ \left. \left.   \frac{1}{2} \bar{M} \dot{Q}^2 + \frac{1}{2}\bar{M}
       \tilde{\Omega}^2 Q^2 -  \tilde{\chi}' x Q    \right)\right] \; ,
 \label{eq:proQ1}
\eea where $\tilde{\omega}^2 = \omega^2 + {i \delta_3}/{(m
\hbar)}$, $\tilde{\Omega}^2 = \Omega^2 + {i \delta_2}/{(M\hbar)}$,
$\bar{M} = M + {i\hbar^2}/{\delta_1}$ and $\tilde{\chi} = \chi +
\delta_4/2$. Making use of the Gaussian approximation, only the
classical path of amide-site ($\bar{Q}$) contributes to the
interaction term \cite{feynmann,kleinert}. It yields, 
\bea
   K(x,x';Q,Q') & = & K_xK_Q \nonumber \\
& = &  \int \cd [x] \exp \left[-\frac{i}{\hbar}\int \d t
\left( \frac{1}{2} m \dot{x}^2
   - \frac{1}{2}\tilde{\omega}^2 x^2  - \tilde{\chi} x \bar{Q} \right) \right] \nonumber\\
   &&\times \int \cd[Q] \exp \left[-\frac{i}{\hbar}\int \d t \left( \frac{1}{2}
\bar{M} \dot{Q}^2
   - \frac{1}{2} \bar{M}\tilde{\Omega}^2 Q^2\right)\right] \; .
    \label{eq:proQ2}
\eea

The propagator $K_Q$ is just a harmonic oscillator, and the solution
is well known  \cite{kleinert,ingold},
\bea
 K(Q,t;Q',0) &=& \exp \left[
 -i\frac{\pi}{2}\left(\frac{1}{2}+ \left|\frac{\tilde{\Omega}
 t}{\pi}\right|\right)\right] \sqrt{\frac{\bar{M}\tilde{\Omega}}{2\pi \hbar
 |\sin(\tilde{\Omega} t)|}} \nonumber\\
 &&\times \exp \left \{ \frac{ i\bar{M} \tilde{\Omega}}{2\hbar
|\sin(\tilde{\Omega} t)|}
 \left[(Q'^2+Q^2)\cos(\tilde{\Omega} t)-2 Q Q' \right] \right \} \; .
 \label{eq:harmonik1}
\eea where $\|x\|$ denotes the largest integer smaller than $x$.

The propagator of $K_x$ is a driven harmonic oscillator and the
solution is also known \cite{kleinert,ingold}. The driven function
$\bar{Q}$ is the classical solution of equation of motion (EOM) of $Q$,
that is a harmonic oscillator. Taking the solution
$\bar{Q}=\bar{Q}_0 \sin(\tilde{\Omega} t)$ and substituting it into the path
integral solution of a driven harmonic oscillator, then
the propagator becomes \cite{kleinert,ingold},
\be
 K(x,t;x',0) =  \exp \left[
-i\frac{\pi}{2}\left(\frac{1}{2}+\left|\frac{\tilde{\omega}
 t}{\pi}\right|\right)\right] \sqrt{\frac{m\tilde{\omega}}{2\pi \hbar
|\sin(\tilde{\omega}
 t)|}}  \ex^{\frac{ i}{\hbar} S_\mathrm{cl}} \; ,
 \label{eq:harmonik2}
\ee
where $S_\mathrm{cl}$ is given by,
\bea
  S_\mathrm{cl} &=& \frac{m\tilde{\omega}}{2 \sin(\tilde{\omega} t)} \left[ (x^2
+ x'^2)
  \cos(\tilde{\omega} t) - 2 x x' \right] \nonumber\\
  &&+ \frac{\tilde{\chi}\bar{Q}_0
x'}{(\tilde{\Omega}^2-\tilde{\omega}^2)\sin(\tilde{\omega} t)}
  \left[\tilde{\omega}\cos(\tilde{\omega}t)\sin(\tilde{\Omega}t)-
  \tilde{\Omega}\sin(\tilde{\omega}t)\cos(\tilde{\Omega}t)
  \right]\nonumber\\
  &&+ \frac{\tilde{\chi}\bar{Q}_0
x}{(\tilde{\Omega}^2-\tilde{\omega}^2)\sin(\tilde{\omega} t)}
     \left[\tilde{\Omega}\sin(\tilde{\omega}t)-\tilde{\omega}\sin(\tilde{\Omega}t)
     \right]
   \label{eq:proX1}\\
 &&+\frac{\tilde{\chi}^2\bar{Q}_0^2}{m\tilde{\omega}(\tilde{\Omega}^2-\tilde{
\omega}^2)\sin(\tilde{\omega} t)}
  \left[ A \cos(\tilde{\omega}t)\sin^2(\tilde{\Omega}t)+B \sin(\tilde{\omega}t)\sin(2\tilde{\Omega}t)
  -\frac{\tilde{\omega}}{4}t\right] \; .
  \nonumber
\eea
with
$A=\tilde{\omega}(\tilde{\omega}\tilde{\Omega}+1)/(4(\tilde{\Omega}^2-\tilde {
\omega}^2))$
and
$B=-\tilde{\omega}(\tilde{\Omega}^2+\tilde{\omega}^2)/(8\tilde{\Omega}(\tilde{\Omega}^2-\tilde{\omega}^2))$.

Combining Eqs. (\ref{eq:harmonik2}) and (\ref{eq:proX1}), the propagator
$K(x,x';Q,Q';t,t')$ becomes,
\bea
K(Q,x,t;Q',x',0) &=& \exp \left[
 -i\pi\left(\frac{1}{2}+\frac{1}{2}\left|\frac{(\tilde{\Omega}+\tilde{\omega})
 t}{\pi}\right|\right)\right]
  \nonumber\\
  &&\times \sqrt{\frac{\bar{M}\tilde{\Omega}}{2\pi\hbar
 |\sin(\tilde{\Omega} t)|}} \sqrt{\frac{m\tilde{\omega}}{2\pi \hbar
|\sin(\tilde{\omega}
 t)|}}
  \label{eq:propagator}\\
 &&\times \exp \left \{ \frac{ \bar{iM} \tilde{\Omega}}{2\hbar
|\sin(\tilde{\Omega} t)|}
 \left[(Q'^2+Q^2)\cos(\tilde{\Omega} t)-2 Q Q' \right]  +\frac{ i}{\hbar}
S_\mathrm{cl}
\right \} \; .
  \nonumber
\eea

\section{Thermodynamic properties}
\label{sec:tp}

The discussion of a system interacting with heat bath is characterized
by the temperature $T$. The state of those systems is therefore given by an
equilibrium density matrix which can be obtained by performing a transformation
$t\rightarrow \tau = -i \hbar \beta$ in the propagator with $\beta=1/(k_B T)$
\cite{kleinert,ingold}. The density matrix is actually the propagator with
$\rho(x,x,Q,Q,\tau,0)=K(Q,x;Q,x,\tau,0)$. Substituting $t\rightarrow
\tau=-i\hbar \beta$ into Eq. (\ref{eq:propagator}), one obtains,
\bea
 \rho_{\beta}(x,Q) &=& \exp \left[
 -i\pi\left(\frac{1}{2}-\frac{i}{2} \left|\frac{(\tilde{\Omega}+\tilde{\omega})
 \hbar \beta }{\pi} \right|\right)\right]\nonumber\\
 &&\times 
\sqrt{\frac{\bar{M}\tilde{\Omega}}{2\pi \hbar
 \sinh(\tilde{\Omega} \hbar \beta)}} \sqrt{\frac{m\tilde{\omega}}{2\pi \hbar \sinh(\tilde{\omega}
 \hbar \beta)}} \nonumber\\
 &&\times \exp \left \{ -\frac{ \bar{M} \tilde{\Omega}}{\hbar}
 \tanh(\frac{1}{2}\Omega \hbar \beta) Q^2 -\frac{ m \tilde{\omega}}{\hbar}
 \tanh(\frac{1}{2}\omega \hbar \beta) x^2 \right . \nonumber\\
  &&+ \left .
\frac{2\tilde{\chi}\bar{Q}_0}{\hbar(\tilde{\Omega}^2-\tilde{\omega}^2)}
  \left[\tilde{\Omega} \sinh^2(\frac{1}{2}\tilde{\Omega} \hbar \beta)-\tilde{\omega}
  \sinh(\tilde{\Omega}\hbar \beta) \sinh(\frac{1}{2}\tilde{\omega}\hbar \beta) \right] x
    \right . \nonumber\\
   &&+ \left .\frac{\tilde{\chi}^2 \bar{Q}_0^2}{m \tilde{\omega}\hbar^2
(\tilde{\Omega}^2-\tilde{\omega}^2)}
   \left[A\Omega \sinh^2(\tilde{\Omega} \hbar \beta)\coth(\tilde{\omega}\hbar \beta)
  + B \sinh(2\tilde{\Omega} \hbar \beta) \right. \right. \nonumber\\
  &&- \left. \left. \frac{\tilde{\omega}\hbar \beta}{4\sinh(\tilde{\omega} \hbar
\beta)}\right]
   \right \} \; .
  \label{eq:densitas}
\eea

Having the density matrix at hand, in the statistical mechanics one can
consider partition function \cite{feynmann},
\be
 Z(\beta) = \ex^{-\beta F} = \int dx \int dQ \, \rho_{\beta}(x,Q) \; .
 \label{eq:partisi1}
\ee
Substituting Eq. (\ref{eq:densitas}) into
Eq. (\ref{eq:partisi1}) and  using the hyperbolic manipulation yield,
\bea
 Z(\beta)&=& \frac{1}{2 \sinh(\frac{1}{2} \tilde{\Omega}\hbar
 \beta)} \frac{1}{2\sinh(\frac{1}{2} \tilde{\omega}\hbar
 \beta)} \,
 \exp \left[ -\frac{i\pi}{2}\left(1- \left|\frac{\tilde{\Omega}
 \hbar \beta}{\pi} \right| + \left| \frac{\tilde{\omega}
  \hbar \beta}{\pi} \right| \right) \right]
    \nonumber\\
   && \times \exp \left[ \frac{\tilde{\chi}^2\bar{Q}_0^2\left\{ \tilde{\Omega}
\sinh^2(\frac{1}{2}\tilde{\Omega} \hbar \beta)-\tilde{\omega}
  \sinh(\tilde{\Omega}\hbar \beta) \sinh(\frac{1}{2}\tilde{\omega}\hbar \beta)
  \right\}^2}{\hbar \bar{M} \tilde{\Omega}(\tilde{\Omega}^2-\tilde{\omega}^2)\tanh(\frac{1}{2}\tilde{\omega \hbar \beta })}
    \right.
  \label{eq:partisi2}\\
  &&+ \left. \frac{\tilde{\chi}^2 \bar{Q}_0^2 \left \{A\Omega
\sinh^2(\tilde{\Omega} \hbar \beta)\cosh(\tilde{\omega}\hbar \beta)
    + B \sinh(2\tilde{\Omega} \hbar \beta)\sinh(\tilde{\omega} \hbar \beta)
   -  \frac{1}{4}\tilde{\omega}\hbar \beta \right \}}
   {m \tilde{\omega}\hbar^2 (\tilde{\Omega}^2-\tilde{\omega}^2)\sinh(\tilde{\omega} \hbar \beta)}
    \right]
    \; ,  \nonumber
\eea
by borrowing the Gaussian integral, $\int dx \, \exp(-ax^2+bx+c) =
\sqrt{\pi/a} \, \exp[{b^2}/{(4a)} + c]$.

For an open quantum system, such as the Davydov-Scott monomer, the changes of
the surrounding environment entropy must be taken into account. The effect
is conveniently incorporated by the specific heat which is also experimentally
measurable. It describes the quantities of heat that must be added to a system
in order to increase its temperature, and defined as \cite{feynmann},
\be
    C = k_B \beta^2 \frac{\pd^2 \ln Z}{\pd \beta^2}\; .\
     \label{eq:thermo}
\ee
Bringing Eq. (\ref{eq:partisi2}), one gets,
\be
  \frac{C}{k_B} =  \frac{I^2}{\sinh^2I} +  \frac{w^2}{\sinh^2(w)} +
        \tilde{\chi}^2 \Lambda(\beta) \; ,
     \label{eq:cv1}
\ee
where $I = {\tilde{\Omega}\hbar \beta}/2$,
$w={\tilde{\omega}\hbar \beta}/2$ and,
\bea
  \Lambda(\beta) &=&
  \frac{\bar{Q}_0^2\beta^2 }{2\hbar} \frac{\pd^2}{\pd \beta^2}
   \left\{ \frac{\left[ \tilde{\Omega} \sinh^2I - \tilde{\omega}
  \sinh(2I) \sinh w \right]^2}{ \bar{M}
  \tilde{\Omega}(\tilde{\Omega}^2-\tilde{\omega}^2)\tanh w }
     \right. \nonumber\\
     &&+ \left. \frac{  A\Omega \sinh^2(2I)\cosh(2w)
  + B \sinh(4I) \sinh(2w)
  - \frac{1}{2}w }
   {m \tilde{\omega}\hbar (\tilde{\Omega}^2-\tilde{\omega}^2)\sinh(2w)}
    \right\} \; .
  \label{eq:cv20}
\eea
For small coupling case we have approximately $\sinh(x)\sim
x$, $\cosh(x)\sim 1$ and $\tanh(x)\sim x$ to yield,
\be
  \Lambda(\beta) =\frac{ \bar{Q}_0^2}{2\hbar (\tilde{\Omega}^2-\tilde{\omega}^2) }
 \left( \frac{3}{4}\tilde{\Omega}^3-\frac{3}{2}\tilde{\Omega}^2\tilde{\omega}
^2-3\tilde{\omega}^4\tilde{\Omega} \right)
  \beta^3\; .\
  \label{eq:cv3}
\ee
The specific heat in Eq. (\ref{eq:cv1}) can be read as,
\be
  C = C_\mathrm{amide-site} + C_\mathrm{amide-I} + C_\mathrm{mixing} \; .
  \label{eq:cv2}
\ee
If there is no coupling between amide-site and amide-I, \ie $\tilde{\chi}=0$,
this is just the total specific heat of two independent harmonic oscillators.

\begin{figure}[t]
\begin{center}
  \centering
   \includegraphics[width=1\textwidth]{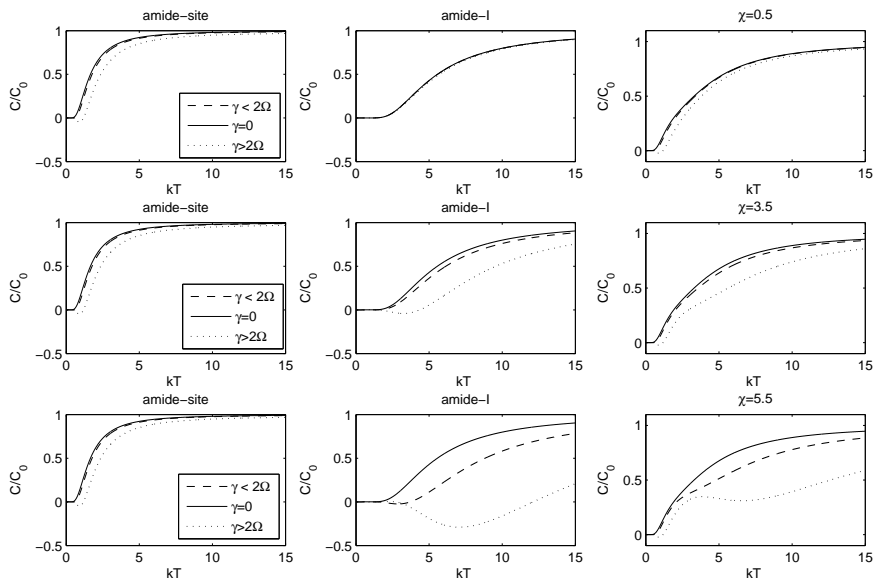}
   \caption{The temperature dependence of normalized specific heat for the first (left), second (middle) and total (right) terms in Eq. (\ref{eq:cv2}) for various values of $\gamma$.}
   \label{fig:heat}
\end{center}
\end{figure}

The result is depicted in Fig. \ref{fig:heat} for three cases corresponding to
the values of $\gamma$, that is $\gamma=0$, $\gamma < \Omega$ and $\gamma >
\Omega$. The case of non-zero but small $\gamma$ has similar behavior as
non-damping case. All of them coincide each other at low temperature and tend
to be asymptotically constant at high temperature. On the other hand, large
$\gamma$ reduces the specific heat for the whole region of temperature. This
means the environtmental effect contributes destructively to the specific heat,
and the system requires less energy to increase the temperature to reach the
equilibrium.

The application of the present model to the $\alpha-$helix protein requires proper knowledge on the coupling constant between amide-I and amide-site \cite{scott}. There are some attempts to determine its allowed range through several methods like the Ab initio calculation and also the extraction from the experimental data as well. The value is found to be within 7 pN and 62 pN \cite{scott}. Considering the amide-site mass $5.7 \times 10^{-25}kg$ and the string constant $\kappa=58.5$ Nm$^{-1}$, one immediately obtains the coupling constant $\chi' = 1.7, 11.6, 18.2$ corresponding to $\chi = 0.5, 2.5, 5.5$ pN. Note that most of previous works takes $\chi = 62$ pN. However, those works do not take into account the thermal bath effect. On the other hand, including the thermal bath as done in the present paper enhance the contribution of amide-I and amide-site interaction.

The result is depicted in Fig. \ref{fig:heat} showing the
temperature dependencies of the normalized specific heat for
various values of $\gamma$. The left, middle and right figures
correspond to the contribution of the first term, the second term
and the total in Eq. (\ref{eq:cv2}) respectively. From the figures one can conclude that 
the results are sensitive to the size of coupling constant $\chi$ at intermediate temperature. 
Especially, the amide-I is more affected than amide-site. The reason is because amide-I has higher frequency than amide-site. Moreover, amide-I is also suppressed significantly by thermal bath contribution which indicates the dependencies of system frequency on the effect of thermal bath  as already pointed out by Ingold \etal  through Caldiora-Lenggets formalism \cite{ingold}.

It should be remarked that at low temperature region large environmental effect
induces an anomaly, that is the specific heat is getting negative. This anomaly
has also been observed by Ingold \etal \cite{ingold2} for free harmonic
oscillator using Caldiora-Lenggets formalism, and by us using full-quantum
approach and the Lindblad formulation of master equation \cite{sulaiman}.

\section{Summary}

The interaction of Davydov-Scott monomer with thermal bath is
investigated using the Lindblad open quantum system formalism. In
contrast with previous work by Cuevas \etal \cite{cuevas}, the
statistical partition function is calculated instead of solving
the EOM itself.

Using path integral one can calculate the propagator of the Lindblad equation.
Under an assumption that the diffusion term is dominant, we have shown that the
environment
contributions shift the kinetic, potential and interaction terms in the
lagrangian. The mixing term is survive only if the frictional damping rate
is taken into account.

In the open quantum system the damping coefficient $\gamma$
represents the relaxation time due to interaction with the
environment. Non-zero $\gamma$ contributes destructively to the
specific heat. The higher value of $\gamma$ corresponds to the
shorter relaxation time, and it induces specific heat anomaly as pointed out in previous works \cite{ingold2,sulaiman}. It is also found that large interaction between amide-I and amide-site is not preferred for a stable Davydov-Scott monomer.

\section*{Acknowledgments}
AS thanks the Group for Theoretical and Computational Physics
LIPI for warm hospitality during the work. This work is funded by the Indonesia
Ministry of Research and Technology and the Riset Kompetitif LIPI in fiscal year
2010 under Contract no.  11.04/SK/KPPI/II/2010. FPZ is supported by Riset KK
2010 Institut Teknologi Bandung.

\bibliographystyle{elsarticle-num}
\bibliography{Sulaiman}

\end{document}